\documentclass[prb,preprint]{revtex4-1} 

\usepackage{amsmath}
\usepackage{amssymb}
\usepackage{amsfonts}
\usepackage{graphicx}

\usepackage{verbatim}

\begin{document}

\title{Relativity free of coordinates}

\author{Serge A. Wagner}
\email{s\_wagner@mail.ru}
\affiliation{
Moscow Institute of Physics and Technology, Dolgoprudny, Moscow Oblast 141701}

\date{\today}

\begin{abstract}
The concept of a physical space, which actualizes Euclidean geometry, is not confined to the statics of solids but extensible to the phenomena where Newtonian mechanics is valid, defining its concept of time. The laws of propagation of electromagnetic disturbances modify Newtonian formalism for sufficiently fast free motions within each spatial domain of its validity for slow motions and introduce the extended concept of time by uniting those of Newtonian which can exist in different spatial domains of their validity.

A boost direction for a pair of physical spaces is that spatial direction in one of the spaces along which the other space moves. Free motions of point particles make an instrumentation for identifying the boost direction as well as events on a straight line along that direction. The concept of a boost direction secures the physics-based formulation of the basic relativity effects, which eventually results in the relation between two arbitrary spaces in terms of their position vectors and time of a given event.

The obtained transformation rules for the components of the position vector differ from the vector-like relationship known in the literature because the latter actually deals with column vectors made of Cartesian coordinates of true vectors and appears identical to a coordinate transformation referred to as a boost.

Within the physics-based approach, the transformation of coordinates implies specifying coordinate systems in each frame in terms of physical objects/directions. This yields a logically consistent and physically meaningful presentation of the coordinate transformations commonly exploited in the special relativity theory, which makes manifestations of those transformations evident. Specifically, for a Cartesian coordinate system subjected to a boost, the coordinate-free technique of reasoning allows one to evaluate its apparent distortion easily.
\end{abstract}

\maketitle

\section{Motivation}\label{motivation}

The safely correct development and credible presentation of a physical theory implies identifying physical phenomena to which the theory is assuredly applicable, however small such a validity domain would seem to be at the outset.

Historically, the special relativity theory came along in attempts to make Newtonian description of moving macroscopic bodies consistent with optical phenomena observable by means of these bodies. Einstein's initial defining paper\cite{Einstein1905} on relativity combines the principle of relativity from Newtonian mechanics with an idea that propagation of light does not depend on the motion of its source and can be used to define simultaneous events at different places. This combination turned out to be a physically reasonable way to arrive at Lorentz transformation, reproduced in subsequent monographs\cite{post-Einsteinian monographs} and textbooks\cite{textbooks}.

The alternative post-Einsteinian presentation\cite{Fock1964} of the relativity theory brought Einstein's attempt to identify the physical foundations of the problem to its actual formal source: the required change of a coordinate system should preserve the form of free motion of point particles while keeping the speed of light \(c\) as a universal limitation on the speed of any particle. Such a mathematical approach proves to be physically important because it provides a consistent description of the macroscopically perceived (commonly referred to as kinematic\cite{Hagedorn1964}) part of experimental particle physics:  Registering devices embedded in the solid wall of an accelerator can embody an inertial reference frame along with its Euclidean geometry; registrable/inferable collisions between particles are events.

Up to the present time, all expositions of the special relativity theory heavily rely on Cartesian coordinates for both formulating premises and developing inferences. Within macroscopic physics, resorting to that handy mathematical technique imposes no restriction on the applicability of the theory, though it comes into apparent conflict with the meaningful practice of exploiting coordinates in physics and engineering. Such a practice implies that the choice of coordinate systems should facilitate the application of basic/general regularities/rules to a particular problem, so that the actual spatial configuration of the problem can determine whether a suitable coordinate system is Cartesian, or orthogonal curvilinear, or even special such as barycentric etc.

In the next section the reader can find a remainder of what physics secures the existence of frames and underlies the validity of Euclidean geometry. Section~\ref{basic effects} presents the concept of a boost direction, which enables the subsequent physics oriented inference of the well known spatiotemporal effects related to the special relativity theory. Section \ref{transformation} exploits these effects to obtain the relation between the time moments and the position vectors of a given event in two frames. This relation entails a logically consistent and physically meaningful presentation of the well known coordinate transformations in Section~\ref{coordinate transformations}. 

\section{Physical spaces}

\subsection{Statics and foundations of Euclidean geometry}\label{statics}

Within macroscopic physics, a physical space can be provisionally viewed as a set of positions of rigidly connected small bodies, where a rigid construction means that the spatial relations between its parts obey the rules of Euclidean geometry. Making use of this mathematical structure in the description of undeformed (or, equivalently, non-accelerating and non-rotating) macroscopic solids is evidently valid, which suggests that the regularities of statics (supplemented with Hooke law to resolve statics' ambiguities) can underlie Euclidean geometry. This mathematical construct is commonly believed to be applicable to a wider range of spatial scales as well as a larger class of physical phenomena. So other regularities, such as those of electromagnetism and gravity, may take part in maintaining Euclidean geometry for the set of all possible positions of sufficiently small interacting bodies.\cite{geometry due to electromagnetism or gravity}

If one selects from the rigidly connected parts some that make up the coordinate system, then one gets a (necessarily inertial\cite{non-inertial frame}) reference frame. Usually a way of measuring time is also associated with an inertial reference frame. Since the concept of time is perceived as independent of coordinates and thus exists for a physical space as a whole, in this article the time variable is associated with a physical space, too.

As of now, the group of motions of rigid bodies (described axiomatically) is the only mathematical structure that reminds us of the physical foundation of Euclidean geometry. The group includes spatial translations \(\hat T\) and rotations \(\hat R\). The additive representation of a spatial translation is usually referred to as a spatial vector. One can use rotations to introduce an angle between two vectors etc. When the orthonormal vectors \(\mathbf{e}_x, \mathbf{e}_y, \mathbf{e}_z\) represent the translations along three mutually perpendicular directions, the decomposition
\begin{equation}\label{position vector decomposition}
\Delta\mathbf{r}=\Delta x\,\mathbf{e}_x+\Delta y\,\mathbf{e}_y+\Delta z\,\mathbf{e}_z
\end{equation}
of a displacement (the change of a position vector \(\mathbf{r}\)) is just what defines Cartesian coordinates.

\subsection{Newtonian dynamics and the applicability extension for Euclidean geometry}\label{geometry}

The previous section refers those interested in the origin of Euclidean geometry to the physics of solids, which may not be a reasonable starting point of an investigation aimed something in the physics of fields and particles.

Meanwhile, it is Newtonian laws that make an established formalism of mechanics. Within the limitations imposed by microscopic phenomena, Newtonian formalism is believed to be applicable to each sufficiently small part of a macroscopic body, referred to as a point particle.

To calculate how the position vector \(\mathbf{r}_a\) of each particle \(a\) is changing over time, a theorist should invoke Newton's second law
\begin{equation}\label{Newton's second Law} 
\left\{ m_a\frac{d^2\mathbf{r}_a}{dt^2}=\sum\limits_b\mathbf{F}_{ab}\right\}.
\end{equation}
Here and hereinafter the notation \(\left\{g_a\right\}\) is used for the list of expressions \(g_a\) where the label \(a\) runs over all its values.

Within the purely mechanical formalism, the force of action of a particle \(b\) on a particle \(a\) cannot be but conservative:
\[
\mathbf{F}_{ab}=-\nabla_a U(|\mathbf{r}_a-\mathbf{r}_b|),
\]
where \(\nabla_a\) denotes the nabla operator designed to act on functions of the position vector \(\mathbf{r}_a\).

For an isolated system of N particles and given functions \(\left\{ U(|\mathbf{r}_a-\mathbf{r}_b|)\right\}\), the decomposition \eqref{position vector decomposition} turns Eqs.~\eqref{Newton's second Law} into a closed set of 3N differential equations of second order. Its general solution
\begin{equation}\label{positions evolution}
\left\{ \mathbf{r}_a=\mathbf{r}_a\left(t+\tau; E, \mathbf{P}, \mathbf{L}, \left\{I_{\mathcal{A}}\right\}\right) \right\}
\end{equation}
involves 6N arbitrary constants, of which one is the time reference shift \(\tau\) while seven can be the total energy \(E\) and the components of the total momentum \(\mathbf{P}\) and the total angular momentum \(\mathbf{L}\). In other words, inversion of the set of Eqs.~\eqref{positions evolution} can yield 6N-8 constants of motions \(\left\{I_{\mathcal{A}}=I_{\mathcal{A}}\left(\left\{\mathbf{r}_a\right\},\left\{\dot{\mathbf{r}}_a\right\}, t\right) \right\}\), possibly specific for each sufficiently small interval of \(t\), in addition to the seven universal constants of motion:
\begin{equation*}
E=\frac12\sum\limits_am_a\dot{\mathbf{r}}_a^2 + \sum\limits_{a\ne b}U(|\mathbf{r}_b-\mathbf{r}_a|),
\end{equation*}
\begin{equation*}
\mathbf{P}=\sum\limits_a m_a\dot{\mathbf{r}}_a,
\end{equation*}
\begin{equation*}
\mathbf{L}=\sum\limits_a m_a[\mathbf{r}_a\times\dot{\mathbf{r}}_a].
\end{equation*}

Of these functions, only \(\left\{I_{\mathcal{A}}\right\}\) essentially represent the internal/relative motions of the particles while \(E\) describes the overall intensity of the whole motion, which can be arbitrary normalized by the choice of the unit of the time \(t\); \(\mathbf{P}\) and \(\mathbf{L}\) correspond to the well known global directions related to the whole motion.

The formalism based on the use of position vectors, their decompositions \eqref{position vector decomposition} and Eqs.~\eqref{Newton's second Law} may not be a consistent introduction to foundations since it refers to the regularities/notions left without description in terms of the physics phenomena involved. To build a foundational construction straightforwardly, one should identify some real or possible stationary objects with Euclidean points so that the appropriate constants of motion can approximate the values of the one-point vector field \(\left\{\mathbf{e}_{\alpha}\right\}\) in Eq.~\eqref{position vector decomposition} , two-point scalar field known as distance (Euclidean length) etc.

The well-known manifestations of quantization preclude implementing such a construction at sufficiently small scales.\cite{quantization}  As a result, nor Eqs.~\eqref{Newton's second Law} nor Eqs.~\eqref{positions evolution} appear applicable at those scales. Nevertheless, to construct Euclidean geometry (or, at least, some pregeometry) one could still adopt the motions of the interacting particles as primitive notions described with a set of experimentally identifiable relations between them. The formulation of such relations (which would then make a low-level automatics of mechanics) is far from the goals of this text, but there is hardly any doubt that the set of motions of interacting particles is sufficiently rich to support Euclidean geometry and, therefore, the concept of a space. At this logical level, the relations that underlie Euclidean geometry are not separable from those that eventually give rise to the existence of the constants of motion.

\subsection{Physical spaces and relativity principle}\label{frames in relativity theory}

As discussed in the previous section, within the familiar high-level formalism the concept of a physical space manifests itself by means of a position vector \(\mathbf{r}\), Euclidean nature of which is most likely secured by non-relativistic classical physics. At least one connection between physical spaces and regularities of classical physics reveals itself in terms of a position vector: Eqs.~\eqref{Newton's second Law} keep their form when one changes a space B for a space A in accordance with the transformation rule
\begin{equation}\label{Galilean transformation}
\left.
\begin{aligned}
&\mathbf{r}^{(\mathrm{B})}=\mathbf{r}^{(\mathrm{A})} +\mathbf{a} -\mathbf{v}^{(\mathrm{A})}_{\mathrm{B}}t^{(\mathrm{A})},\\
&t^{(B)} =t^{(A)} +\tau;
\end{aligned}
\right\}
\end{equation}
where \(\mathbf{a}\), \(\mathbf{v}^{(\mathrm{A})}_{\mathrm{B}}\) and \(\tau\) are given parameters. Here and hereinafter the superscript (F) indicates that a quantity \(q^{(\mathrm{F})} \) is initially defined in a space F. (But as far as the transformation \eqref{Galilean transformation} is valid between any pair of spaces, one can actually define a position vector \(\mathbf{r}\) in any space.)

From the early days of the relativity theory, the following generalization of the above statement is regarded as a more or less universal principle, called the principle of relativity: The mutual disposition and the relative translational uniform rectilinear motion of two spaces cannot manifest itself in the description of physical phenomena within one of these spaces. Equivalently, physical laws have the same formulations in different (necessarily non-accelerated non-rotating) spaces.\cite{first formulation of the relativity principle}

In the context of Newtonian mechanics, the meaning of the relativity principle is plain: translational uniform rectilinear motion of a physical system as a whole with respect to some external (reference) bodies does not affect the motion of the internal parts of the system with respect to each other. However, this idea presumes a partition of the physical system that cannot be seamlessly extended to include electromagnetic phenomena since the decomposition of an electromagnetic field into non-interfering components is possible only without electric charges. This apparent gap is accompanied (and aggravated) by the fact that the transformation \eqref{Galilean transformation} does not preserve the form of source-free Maxwell's equations.

Naturally, an attempt to derive the general transformation that keeps the full formalism of Maxwellian electrodynamics would lead a researcher to a complicated problem. So the developers of the special relativity theory cannot but begin with the simple coordinate transformation called Lorentz transformation (which, in accordance with the purpose  and logic of this article, is explicitly reproduced as late as in Section \ref{Lorentz transformation}.)

Looking into early presentations of the special relativity theory, one can identify the following premise for the formal derivation of Lorentz transformation: the law of motion of a free particle and the laws of propagation of a free electromagnetic field have the same form in all frames.\cite{early presentations} To be exact, in a space F the position \(\mathbf{r}^{(\mathrm{F})}_a\) of a free particle is changing along with the time \(t^{(\mathrm{F})}\) as
\begin{equation}\label{free particle motion}
\mathbf{r}^{(\mathrm{F})}_a=
\mathbf{r}^{(\mathrm{F})}_{a0}+\mathbf{v}^{(\mathrm{F})}_at^{(\mathrm{F})}
\end{equation}
while the positions \(\mathbf{r}^{(\mathrm{F})}\) taken by an electromagnetic wavefront (wave phase-front) at the time \(t^{(\mathrm{F})}\) from a point source flashed at the position \(\mathbf{r}^{(\mathrm{F})}_0\) at the time \(t^{(\mathrm{F})}_0\) satisfy
\begin{equation}\label{spherical wavefront equation}
\left(\mathbf{r}^{(\mathrm{F})}-\mathbf{r}^{(\mathrm{F})}_0\right)^2=c^2\left(t^{(\mathrm{F})}-t^{(\mathrm{F})}_0\right)^2.
\end{equation}
The propagation speed \(c\) of an electromagnetic spherical wavefront is the same in any space.
As long as the goals of one's inference are limited by the derivation of rules equivalent to those of Lorentz transformation, one can confine oneself with the limiting form of Eq.~\eqref{spherical wavefront equation} for an infinitely far source position, i.e. the equation
\begin{equation}\label{plane wavefront equation}
\left(\mathbf{n}^{(\mathrm{F})}\cdot\mathbf{r}^{(\mathrm{F})}\right) - ct^{(\mathrm{F})}=\mathrm{const}
\end{equation}
for a plane wavefront which propagates in the direction of the unit vector \(\mathbf{n}^{(\mathrm{F})}\).

One can view the above statements as some realization of Einstein's original intention to extend the principle of relativity to the propagation of electromagnetic waves, but without his explicit redefinition of the time variable. However, the next generation of authors has dispensed with both Maxwell equations and Eq.~\eqref{spherical wavefront equation} in their introductions to the relativity theory. In the post-Einsteinian derivations of the transformation rules between two frames one finds the principle of relativity replaced by the requirement to preserve the form of Eq.~\eqref{free particle motion} supplemented with the condition
\begin{equation}\label{light ray velocity condition}
\left(\mathbf{v}^{(\mathrm{F})}\cdot\mathbf{v}^{(\mathrm{F})}\right)=c^2
\end{equation}
to include "the motion of a light signal" (in effect, the propagation of the intersection point of an electromagnetic plane wavefront with one of its associated ray paths.)\cite{Fock presentation} As a result, the physics that underlies the special relativity theory has been reduced to that of free point particles with the universal limitation on their speed, equal to \(c\) in any space.

The use of free particles in a reasoning is evidently restricted by the processes of particles' interaction. But as long as one neglects the extension and the duration of those processes, one can exploit an interaction act as a representation of a basic identifiable entity usually referred to as an event. In other words, each event appears to be real or possible interception of two (or more) free particles. Then one can exploit experimentally identifiable relations between motions of free particles to establish relations between events. If need, the interaction between particles can be assumed so small as not to change their motion. (In general, one need exploiting Eq.~\eqref{free particle motion} for the motion of each particle \(a\) between the points of interception, where the parameters \(\mathbf{v}^{(\mathrm{F})}_a\) may change.) This technique along with the basic appliance of the relativity principle is a main tool used in the next section.

Here it is also worth noting that the free motion of particles and the propagation of light rays are not sufficient to construct Euclidean geometry, so the attempts to extend its validity to arbitrarily fast processes could end up with nothing but a new postulate.\cite{relativistic construction of geometry} In order to have material carriers of Euclidean geometry, presentations of the relativity theory have no choice but to borrow the notion of space in Newtonian mechanics and then introduce many spaces uniformly moving relative to each other.

When someone applies Newton's second law \eqref{Newton's second Law} to a physical system which consists of weakly interacting (e.g., widely separated) parts, he might think that Newtonian mechanics should involve some means to identify motions in such parts as simultaneous processes. Actually, Eqs.~\eqref{Newton's second Law} are well known to have originated from the regularities \label{position evolution} revealed by experiments/observations related to strongly interacting physical bodies, especially gravitationally bound ones, such as Sun and planets. But in such a system, the existence of the time \(t\) is an inherent property of its (almost periodic or quasi-periodic) motion. If \(v\) is a characteristic speed of such motion, then for a given timescale \(\Delta t\) it secures synchronization in changing physical quantities over the region of size \(l\sim v\Delta t\). However, the special relativity theory implies the speed of light \(c\) as a characteristic speed and, therefore, considers the region of size \(L\sim c\Delta t\gg l\). In other words, within the special relativity theory, Newtonian mechanics can secure the synchronization two processes ("clocks") but only at one spatial point.

Exploiting Eq.~\eqref{spherical wavefront equation}/Eq.~\eqref{plane wavefront equation} or the light rays only (without specifying geometrical structures prematurely) one can synchronize events happened to particles at different positions (as far as one neglects the time delay and the position shift due to interaction of a charged particle and electromagnetic field.) Adopting the synchronization properties formulated in Ref.~\onlinecite[S.~894]{Einstein1905} turns the time \(t\) into a global variable, similar to a spatial coordinate.

In principle, within sufficiently small timescales, Newtonian mechanics can maintain Euclidean geometry only in the vicinity of each event, which means that the geometry of the subset \(t=\,\)const, identified in Section~\ref{statics} also as a space, might be Riemannian.\cite{Fock and geometry} In this text, the global geometry of particles' positions within each extended space is still assumed Euclidean, since it is appropriate for usual practical applications of the special relativity theory as well as common teaching curricula.

\section{Basic effects of the relativity theory}\label{basic effects}

\subsection{Boost direction}\label{boost direction}

For any pair of spaces A and B, there are the velocity \(\mathbf{v}^{(\mathrm{B})}_{\mathrm{A}}\) of A with respect to B and, vice versa, the velocity \(\mathbf{v}^{(\mathrm{A})}_{\mathrm{B}}\). Since each of these vectors is defined as a spatial object in its own space, there can be no procedure of comparing them directly on the basis of Newtonian mechanics. Nevertheless, since all spaces are supposed to be identical in their essential internal properties, one should accept
\begin{equation}\label{velocity symmetry}
|\mathbf{v}^{(\mathrm{B})}_{\mathrm{A}}|=|\mathbf{v}^{(\mathrm{A})}_{\mathrm{B}}|\equiv v_{\mathrm{AB}},\equiv v\quad \gamma_{\mathrm{AB}}\equiv\gamma\left(v_{\mathrm{AB}}\right)= \gamma(v)\equiv\gamma
\end{equation}
due to the symmetry of exchanging A and B. (This should be also considered as a part of establishing universal time unit since one cannot be sure that a non-relativistic standard clocks keeps its rate when set in fast motion.)

Let \(\mathcal{LM}\) denote a process that starts with an event \(\mathcal{M}\) and ends with an event \(\mathcal{L}\), and let a number \(t^{(\mathrm{F})}(\mathcal{LM})\) denote the elapsed time in an arbitrary space F, so that
\begin{equation}\label{division of time interval}
t^{(\mathrm{F})}(\mathcal{LM}\cap\mathcal{MN})=t^{(\mathrm{F})}(\mathcal{LM})+t^{(\mathrm{F})}(\mathcal{MN})
\end{equation}
and \(t^{(\mathrm{F})}(\mathcal{LM})=-t^{(\mathrm{F})}(\mathcal{ML})\).

If in the space A one has \(t^{(\mathrm{A})}(\mathcal{LM})=t^{(\mathrm{A})}(\mathcal{RS})\) for arbitrary events \(\mathcal{L}\), \(\mathcal{M}\), \(\mathcal{R}\), \(\mathcal{S}\), then in the space B one finds the same equality \(t^{(\mathrm{B})}(\mathcal{LM})=t^{(\mathrm{B})}(\mathcal{RS})\) because the contrary would allow one to judge about the motion of B with respect to A on the basis of internal data in B, in contradiction with the principle of relativity. Further generalization is possible if one exploits Eq.~\eqref{division of time interval} to partition a process into a series of shorter processes over even intervals of time: \(t^{(\mathrm{A})}(\mathcal{LM})=\sigma t^{(\mathrm{A})}(\mathcal{RS})\) entails \(t^{(\mathrm{B})}(\mathcal{LM})=\sigma t^{(\mathrm{B})}(\mathcal{RS})\) for any natural, rational and, finally, real \(\sigma\). It follows that for any space F the relation \(t^{(\mathrm{F})}(\mathcal{LM})/t^{(\mathrm{F})}(\mathcal{RS})\) shows no dependence on the space while for any process \(\mathcal{LM}\) the relation \(t^{(\mathrm{B})}(\mathcal{LM})/t^{(\mathrm{A})}(\mathcal{LM})\) depends only on \(v\) defined by Eq.~\eqref{velocity symmetry}.

Let a point body \(a\) resting in the space A and a point body \(b\) resting in the space B meet each other at the event  \(\mathcal{O}\) and let another freely moving point body \(g\) meet \(a\) and \(b\) at the events \(\mathcal{A}\) and \(\mathcal{B}\), respectively. By making use of the laws \eqref{free particle motion} of free motion, one can get an expression
\begin{equation}\label{three meetings}
\mathbf{v}^{(\mathrm{F})}_g=\alpha\mathbf{v}^{(\mathrm{F})}_{\mathrm{A}}+(1-\alpha)\mathbf{v}^{(\mathrm{F})}_{\mathrm{B}}
\end{equation}
for the velocity \(\mathbf{v}^{(\mathrm{F})}_g\) of the body \(g\) in the space F where
 \[\alpha=\frac{t^{(\mathrm{F})}(\mathcal{AO})}{t^{(\mathrm{F})}(\mathcal{AO})-t^{(\mathrm{F})}(\mathcal{BO})}\]
is actually independent of F in accordance with the previous analysis.

Eq.~\eqref{three meetings} allows one to define a one-parametric family of spaces \(\mathrm{G}[\alpha]\) where \(g\) is at rest while the bodies \(a\) and \(b\) move along the common straight line in opposite directions. Hereinafter this line is referred to as a boost line of a space \(\mathrm{G}[\alpha]\) for a given \(\alpha\), and the whole family \(\mathrm{G}[\alpha]\) is referred to as a helicoboost class of spaces, specified by its two members A and B.

If one considers two different values \(\alpha=\alpha_1\) and \(\alpha=\alpha_2\) along with two corresponding bodies \(g_1\) and \(g_2\) and applies Eq.~\eqref{three meetings} to the motion of the body \(g_2\) in the space \(\mathrm{G}[\alpha_1]\), one can find that \(g_2\) moves just along the boost line of \(\mathrm{G}[\alpha_1]\). Since \(\mathrm{G}[\alpha_1]\) is an arbitrary member of the helicoboost class and \(g_2\) can represent an arbitrary point body (or even a light signal) that moves along the boost line in \(\mathrm{G}[\alpha_1]\), the boost lines in different spaces of one helicoboost class prove equivalent in a sense: Any point bodies that stay (moving or resting) on the boost line in one space of a given helicoboost class remain (moving or resting) on the boost line in another space of the same helicoboost class.

If, in addition to the bodies \(a\) and \(b\), one chooses more representatives of the spaces A and B, one gets more parallel boost lines. To avoid specifying particular boost lines in a reasoning, one can invoke the direction of \(\mathbf{v}^{(\mathrm{G})}_{\mathrm{B}}\) or \(-\mathbf{v}^{(\mathrm{G})}_{\mathrm{A}}\), which represents a bundle of parallel boost lines in each member G of the helicoboost class. Hereinafter this direction is referred to as a boost direction.

It is possible to define both a boost direction and a helicoboost class of spaces without prior reference to its two members: If there are a space where a set of free point bodies have their velocities parallel or anti-parallel to each other, then there are other spaces where velocities of the these bodies are also parallel or anti-parallel to each other; an equivalence class of such spaces is a helicoboost class; the direction of motion of such free bodies in each space of the helicoboost class is a boost direction.

\subsection{Spatial transverse effect}

The motion of light signals along the boost direction in one space of a given helicoboost class is an important limiting case of the motions considered in the previous section. Since the light signals represent the propagation of electromagnetic plane wavefronts \eqref{plane wavefront equation}, one can conclude that they propagate along the boost direction in any space of the helicoboost class. It follows that simultaneous events in a plane perpendicular to the boost direction in one space appear simultaneous in any other space of the same helicoboost class, where they also occupy a plane perpendicular to the boost direction.

Let the locations of the above simultaneous events make a certain instantaneous arrangement over the plane wavefront. To represent it as a stationary geometric configuration, one should consider intersections some bundle of boost lines with a dense series of parallel wavefronts which propagate along the boost direction. Since such wavefronts and the boost lines are observable in any space of the helicoboost class, so is the stationary planar configuration they generate. The possibility of the common geometric configuration shows that the observers in different spaces can come to agreement with each other about the orientation of their spaces around the boost direction or, in other words, to another equivalence relation between two spaces. This relation allows one to partition the helicoboost class into subclasses of identically oriented spaces. In the following, a subclass of this kind is referred to as a boost class.

To present the above equivalence relation in terms of a relative position vector, one can write:
\begin{equation}\label{transversal plane relation}
\left(\Delta\mathbf{r}\right)^{(\mathrm{A})}_{\perp}\backsim\left(\Delta\mathbf{r}\right)^{(\mathrm{B})}_{\perp}.
\end{equation}
Here and in the rest of the paper, the notation
\begin{equation}\label{decomposition}
\left.
\begin{aligned}
\mathbf{q}^{(\mathrm{A})}&=q^{(\mathrm{A})}_{\parallel}\frac{\mathbf{v}^{(\mathrm{A})}_{\mathrm{B}}}{v}+\mathbf{q}^{(\mathrm{A})}_{\perp},\quad\quad q^{(\mathrm{A})}_{\parallel}\equiv\frac{(\mathbf{q}^{(\mathrm{A})}\cdot\mathbf{v}^{(\mathrm{A})}_{\mathrm{B}})}{v}, \\
\mathbf{q}^{(\mathrm{B})}&=-q^{(\mathrm{B})}_{\parallel}\frac{\mathbf{v}^{(\mathrm{B})}_{\mathrm{A}}}{v}+\mathbf{q}^{(\mathrm{B})}_{\perp},\quad -q^{(\mathrm{B})}_{\parallel}\equiv\frac{(\mathbf{q}^{(\mathrm{B})}\cdot\mathbf{v}^{(\mathrm{B})}_{\mathrm{A}})}{v}
\end{aligned}
\right\}
\end{equation}
describes the decompositions of spatial vectors \(\mathbf{q}^{(\mathrm{A})}\) and \(\mathbf{q}^{(\mathrm{B})}\) in their respective spaces.

Since the symbol \(\left(\Delta\mathbf{r}\right)^{(\mathrm{A})}_{\perp}\) in Eq.~\eqref{transversal plane relation}
denotes an arbitrary relative position vector, which connects two arbitrary points in a plane perpendicular to \(\mathbf{v}^{(\mathrm{A})}_{\mathrm{B}}\), one can view Eq.~\eqref{transversal plane relation} as a notation for mapping a geometric configuration in the space A to that in the space B. Since the principle of relativity does not allow differences in the geometric properties of these configurations, then
\begin{equation}\label{addition property}
\mathbf{f}^{(\mathrm{A})}_1\backsim\mathbf{g}^{(\mathrm{B})}_1\text{ and }\mathbf{f}^{(\mathrm{A})}_2\backsim\mathbf{g}^{(\mathrm{B})}_2\text{ entail } \mathbf{f}^{(\mathrm{A})}_1+\mathbf{f}^{(\mathrm{A})}_2\backsim\mathbf{g}^{(\mathrm{B})}_1+\mathbf{g}^{(\mathrm{B})}_2
\end{equation}
and
\begin{equation}\label{dot property}
\mathbf{f}^{(\mathrm{A})}_3\backsim\mathbf{g}^{(\mathrm{B})}_3\text{ and }\mathbf{f}^{(\mathrm{A})}_4\backsim\mathbf{g}^{(\mathrm{B})}_4\text{ entail } \left(\mathbf{f}^{(\mathrm{A})}_3\cdot\mathbf{f}^{(\mathrm{A})}_4\right)= \left(\mathbf{g}^{(\mathrm{B})}_3\cdot\mathbf{g}^{(\mathrm{B})}_4\right)
\end{equation}
for any spatial vectors \(\mathbf{f}^{(\mathrm{A})}_i\) in the space A and their counterparts \(\mathbf{g}^{(\mathrm{B})}_i\) in the space B.

\subsection{Spatial longitudinal effect}

Suppose in the space A point bodies P, Q, R, S\dots are moving with the same velocity \(\mathbf{v}^{(\mathrm{A})}_{\mathrm{B}}\) while momentarily (detected as) arranged in a straight line along the boost direction, i.e. parallel to \(\mathbf{v}^{(\mathrm{A})}_{\mathrm{B}}\). In the space B these point bodies are at rest in a straight line along the boost direction. If in the space A one has \(l^{(\mathrm{A})}_{\mathrm{B}}(\mathrm{PQ})=l^{(\mathrm{A})}_{\mathrm{B}}(\mathrm{RS})\), then in the space B one finds the correspondent equality \(l_{\mathrm{B}}(\mathrm{PQ})=l_{\mathrm{B}}(\mathrm{RS})\) due to the relativity principle. Here the notation \(l^{(\mathrm{A})}_{\mathrm{B}}(\mathrm{GH})\) refers to a distance between moving bodies G and H momentarily observed in a space A while \(l_{\mathrm{B}}(\mathrm{GH})\) denotes a distance between stationary bodies G and H in a space B. If one exploits partitioning a line segment in a manner similar to the division \eqref{division of time interval} of a time interval, one can eventually come to the similar conclusion that \(l^{(\mathrm{A})}_{\mathrm{B}}(\mathrm{PQ})=\sigma l^{(\mathrm{A})}_{\mathrm{B}}(\mathrm{RS})\) entails \(l_{\mathrm{B}}(\mathrm{PQ})=\sigma l_{\mathrm{B}}(\mathrm{RS})\) for any natural, rational and, finally, real \(\sigma\). It follows that for any space F of the boost class the relation \(l^{(\mathrm{F})}_{\mathrm{B}}(\mathrm{PQ})/l^{(\mathrm{F})}_{\mathrm{B}}(\mathrm{RS})\) shows no dependence on the space while for any pair of bodies P and Q stationary in the space B the relation \(l_{\mathrm{B}}(\mathrm{PQ})/l^{(\mathrm{A})}_{\mathrm{B}}(\mathrm{PQ})\) depends only on \(v\).

The application of the above result is not bounded by a comparison of the distances between two bodies in the two spaces. In fact, the distance \(l^{(\mathrm{A})}_{\mathrm{B}}\) between two moving point bodies in the space A is a distance between two \textit{simultaneous} events of detecting these bodies in the space A. In the space B, these events are not necessarily simultaneous but happening to the same bodies. So the distance \(l_{\mathrm{B}}\) between the bodies is a distance between the events, too. Thus, the relation
\begin{equation}\label{contraction factor}
\frac{l_{\mathrm{B}}}{l^{(\mathrm{A})}_{\mathrm{B}}}=K(v)
\end{equation}
is the same for any pair of events in the straight line along the boost direction, provided that they are simultaneous in the space A.

Let simultaneous elementary events in the space A occur over the length \(l_{\mathrm{A}}\) in a straight line along the boost direction, i.e. along the direction of motion of the space B.  Physically, they together may be an act of detecting a rigid rod embedded into the space A. If a flashlight occurs in the middle of the rod, it takes the same time interval \(t_{\mathrm{A}}=l_{\mathrm{A}}/2c\) for that light to get to each end of the rod. When instantaneously observed in the space B, the length of the rod appears to be \(l_{\mathrm{A}}^{(\mathrm{B})}\). The arrivals of the above mentioned flashlight at the ends of the moving rod are not simultaneous in the space B: along the direction of the rod's motion the time difference makes
\begin{equation}\label{time spread}
\Delta t^{(\mathrm{B})}=\frac{l_{\mathrm{A}}^{(\mathrm{B})}/2}{c+v}-\frac{l_{\mathrm{A}}^{(\mathrm{B})}/2}{c-v}=-\gamma^2\frac{vl_{\mathrm{A}}^{(\mathrm{B})}}{c^2}
\end{equation}
while the distance between these events is
\begin{equation}\label{spatial extension}
l_{\mathrm{B}}=c\frac{l_{\mathrm{A}}^{(\mathrm{B})}/2}{c+v}+c\frac{l_{\mathrm{A}}^{(\mathrm{B})}/2}{c-v}=\gamma^2l_{\mathrm{A}}^{(\mathrm{B})}
\end{equation}
Here and hereafter \(\gamma\equiv\gamma(v)\equiv (1-v^2/c^2)^{-1/2}\).

The equations \eqref{time spread} and \eqref{spatial extension} take account of the possibility that in different spaces a set of the same events may occupy segments of different sizes along their boost direction. Considering the events involved in the inference and the formulation of Eqs. \eqref{time spread} and \eqref{spatial extension} one can find that \(l_{\mathrm{A}}=K(v)l_{\mathrm{A}}^{(\mathrm{B})}\) and \(l_{\mathrm{B}}=K(v)l_{\mathrm{A}}\) in accordance with Eq.~\eqref{contraction factor}. Then Eq.~\eqref{spatial extension} entails \(K(v)=\gamma\), i.e. the length contraction effect
\begin{equation}\label{length contraction}
l_{\mathrm{A}}=\gamma l_{\mathrm{A}}^{(\mathrm{B})}
\end{equation}

\subsection{Time dilation effect}

To get to another well known effect, one can exploit such thought construct as the light clock\cite{light clock}, where a light pulse propagates back and forth between two mirrors held parallel and apart at the fixed distance \(l_0\) in their rest space. In that space, the line of the light propagation is perpendicular to the mirrors and can be referred to as the axis of the clock. The proper duration of the clock's cycle, i.e. the round-trip time of the light pulse in the clock's rest space, makes
\begin{equation}\label{proper period}
\Delta\tau=\frac{2l_0}{c}.
\end{equation}

Let an observer be moving with a speed \(v\) in the light clock's rest space and along the axis of the light clock. The observer can detect the contracted length \(l=l_0/\gamma\) of the light clock and find that the clock's cycle takes
\begin{equation}\label{dilated period}
\Delta t=\frac{l}{c-v}+\frac{l}{c+v}=2\gamma^2 l/c=2\gamma l_0/c.
\end{equation}

Here it is worth remarking that the light clock is an auxiliary dedicated thought construct, which can in no way include the actual reflections of an electromagnetic wave from substance of the mirrors but embodies the properly collated sequence of events only. This means that Eq.~\eqref{proper period} and Eq.~\eqref{dilated period} together entail the time dilation effect
\begin{equation}\label{time dilation}
\Delta t=\gamma\Delta\tau,
\end{equation}
which gives the laboratory time interval \(\Delta t\) between two events at a moving point during its given proper time interval \(\Delta\tau\).

To show relations between the spatiotemporal effects of the relativity theory in one's practice of teaching, one may also address the following inference. 

\begin{figure}[h]
  \centering
 \includegraphics[width=10cm]{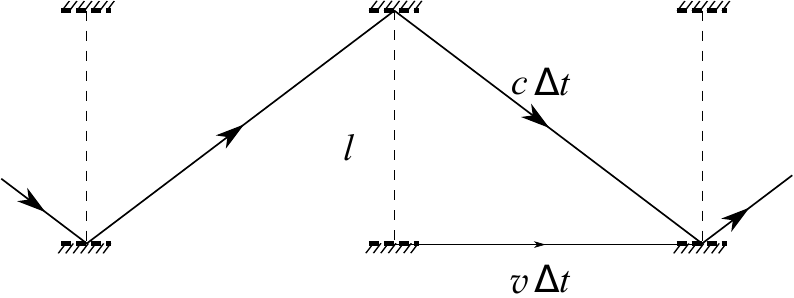}
  \caption{The path of the light pulse in the transversally moving light clock.}
   \label{moving light clock}
\end{figure}

Let the light clock of proper length \(l_0\) rest in the space A so that the axis of the clock is perpendicular to the direction of motion of the space B, i.e. to the boost direction. In the space B, the axis of the clock is also perpendicular to the boost direction, and the main cycle of the clock corresponds to the light signal path shown in Fig.~\ref{moving light clock}. 
Here \(l\) is a length of the clock simultaneously observed in the space B as being in the plane perpendicular to the direction of the clock's motion. In accordance with Fig.~\ref{moving light clock} the duration \(\Delta t\) of the clock's main cycle in the space B should obey the equation
\[l^2+(v\Delta t/2)^2=(c\Delta t/2)^2.\]
This is consistent with the time dilation effect \eqref{time dilation} and the proper duration \eqref{proper period} of the clock's main cycle because the spatial transversal effect \eqref{transversal plane relation} yields \(l=l_0\) when the motion of the light clock and the propagation of the light pulse occur in the same plane.

\subsection{Retardation effect}

Eq.~\eqref{transversal plane relation} suggests that one can generalize Eq.~\eqref{length contraction} and Eq.~\eqref{time spread}: If in the space A simultaneous events occur in the transversal spatial slab \(\Delta r^{(\mathrm{A})}_{\parallel}\) thick, then in the space B they occur within the slab
\begin{equation}\label{A slab contraction}
\Delta r^{(\mathrm{B})}_{\parallel}=\gamma\Delta r^{(\mathrm{A})}_{\parallel}
\end{equation}
and with the time spread
\begin{equation}\label{B slab time spread}
\Delta t^{(\mathrm{B})}=-\gamma\frac{v\Delta r^{(\mathrm{A})}_{\parallel}}{c^2}=-\gamma\frac{(\mathbf{v}^{(\mathrm{A})}_{\mathrm{B}}\cdot\Delta\mathbf{r}^{(\mathrm{A})})}{c^2}.
\end{equation}


\section{Relation between two spaces}\label{transformation}

To describe the relation between the spaces A and B  in terms of the position vectors \(\mathbf{r}^{(\mathrm{A})}\) and \(\mathbf{r}^{(\mathrm{B})}\) and the time moments \(t^{(\mathrm{A})}\) and \(t^{(\mathrm{B})}=0\) of a given arbitrary event \(\mathcal{E}\) one should first of all identify the place of a common reference event \(\mathcal{O}\) with the origins \(\mathbf{r}^{(\mathrm{A})}=0\) and \(\mathbf{r}^{(\mathrm{B})}=0\) of the spaces A and B as well as its time moment with the readings \(t^{(\mathrm{A})}=0\) and \(t^{(\mathrm{B})}=0\) of their clocks.

Let two elementary events occur in the space A at a given time moment \(t^{(\mathrm{A})}\): an event \(\mathcal{O}^{\mathrm{A}}\) occurs at the origin \(\mathbf{r}^{(\mathrm{A})}=0\) while \(\mathcal{E}\) occurs at \(\mathbf{r}^{(\mathrm{A})}\ne 0\). Due to the time dilation effect, the moment of observing \(\mathcal{O}^{\mathrm{A}}\) in the space B equals to the reading \(\gamma t^{(\mathrm{A})}\) of a clock fixed at the origin  \(\mathbf{r}^{(\mathrm{B})}=0\). Due to the time spread effect \eqref{B slab time spread}, in the space B the difference in time between \(\mathcal{E}\) and \(\mathcal{O}^{\mathrm{A}}\)
\[\Delta t^{(\mathrm{B})}=-\gamma\frac{vr^{(\mathrm{A})}_{\parallel}}{c^2}.\]
Therefore, the time moment of observing \(\mathcal{E}\) in the space B
\begin{equation}\label{time transformation}
t^{(\mathrm{B})}=\gamma t^{(\mathrm{A})}-\gamma\frac{vr^{(\mathrm{A})}_{\parallel}}{c^2}. 
\end{equation}

Further, if in the space B the plane \(\Pi_{\mathrm{B}}\) is perpendicular to \(\mathbf{v}^{(\mathrm{A})}_{\mathrm{B}}\) and keeps passing through the origin \(\mathbf{r}^{(\mathrm{B})}=0\), then, in accordance with the analysis in Section \ref{boost direction}, in the space A it forms the plane \(\Pi^{(\mathrm{A})}_{\mathrm{B}}\) perpendicular to \(\mathbf{v}^{(\mathrm{B})}_{\mathrm{A}}\) and moving with this velocity. Evidently, the distance from the event \(\mathcal{E}\) to the plane \(\Pi^{(\mathrm{A})}_{\mathrm{B}}\) makes \(\Delta r^{(\mathrm{A})}_{\parallel}=r^{(\mathrm{A})}_{\parallel}-vt^{(\mathrm{A})}\). If \(\mathbf{r}^{(\mathrm{B})}\) denotes the position of \(\mathcal{E}\) in the space B, then the length contraction effect \eqref{A slab contraction} helps us conclude that
in the space B the distance from the event \(\mathcal{E}\) to the plane \(\Pi_{\mathrm{B}}\) is
\begin{equation}\label{longitude transformation}
r^{(\mathrm{B})}_{\parallel}=\gamma\left(r^{(\mathrm{A})}_{\parallel}-vt^{(\mathrm{A})}\right)
\end{equation}
while Eq.~\eqref{transversal plane relation} yields
\begin{equation}\label{transversal plane transformation}
\mathbf{r}^{(\mathrm{A})}_{\perp}\backsim\mathbf{r}^{(\mathrm{B})}_{\perp}.
\end{equation}

Within the described coordinate-free approach, the relationships \eqref {time transformation}, \eqref{longitude transformation} and \eqref{transversal plane transformation} turn out to be transformation rules between spaces. To show the dependence on the spaces A and B explicitly one should only decipher the notation introduced by Eq.~\eqref{velocity symmetry} and Eq.~\eqref{decomposition}:
\begin{equation}\label{time relation AB}
t^{(\mathrm{B})}=\gamma_{ \mathrm{AB}}\left[t^{(\mathrm{A})}-\frac{(\mathbf{v}^{(\mathrm{A})}_{\mathrm{B}}\cdot\mathbf{r}^{(\mathrm{A})})}{c^2}\right],
\end{equation}
\begin{equation}\label{longitudinal relation AB} 
-\frac{ \left(\mathbf{r}^{(\mathrm{B})}\cdot\mathbf{v}^{(\mathrm{B})}_{\mathrm{A}}\right) }{ v_{\mathrm{AB}} }=
\gamma _{\mathrm{AB}}\left[\frac{\left(\mathbf{r}^{(\mathrm{A})}\cdot\mathbf{v}^{(\mathrm{A})}_{\mathrm{B}}\right)}{v_{ \mathrm{AB}} }-v_{ \mathrm{AB}}t^{(\mathrm{A})}\right],
\end{equation}
\begin{equation}\label{transversal relation AB} 
\mathbf{r}^{(\mathrm{B})}-\frac{\left(\mathbf{r}^{(\mathrm{B})}\cdot\mathbf{v}^{(\mathrm{B})}_{\mathrm{A}}\right)\mathbf{v}^{(\mathrm{B})}_{\mathrm{A}}}{v_{\mathrm{AB}}^2}
\backsim
\mathbf{r}^{(\mathrm{A})}-\frac{\left(\mathbf{r}^{(\mathrm{A})}\cdot\mathbf{v}^{(\mathrm{A})}_{\mathrm{B}}\right)\mathbf{v}^{(\mathrm{A})}_{\mathrm{B}}}{v_{\mathrm{AB}}^2}.
\end{equation}

In a more compact form, the transformation rules \eqref{time relation AB}-\eqref{transversal relation AB} can also be presented as
the mapping
\begin{equation}\label{tr relation}
\begin{pmatrix}
ct^{(\mathrm{B})}\\
\mathbf{r}^{(\mathrm{B})}
\end{pmatrix}
\leftrightarrow
\mathbf{M}^{(\mathrm{B})}_{(\mathrm{A})}\odot\begin{pmatrix}
ct^{(\mathrm{A})}\\
\mathbf{r}^{(\mathrm{A})}
\end{pmatrix}
\end{equation}
where
\begin{equation}\label{vector transformation matrix}
\mathbf{M}^{(\mathrm{B})}_{(\mathrm{A})}\equiv\begin{pmatrix}
\gamma_{\mathrm{AB}} &-\gamma_{\mathrm{AB}}\mathbf{v}^{(\mathrm{A})}_{\mathrm{B}}/c\\
\gamma_{\mathrm{AB}}\mathbf{v}^{(\mathrm{B})}_{\mathrm{A}}/c&1-\gamma_{\mathrm{AB}} \mathbf{v}^{(\mathrm{B})}_{\mathrm{A}}\otimes\mathbf{v}^{(\mathrm{A})}_{\mathrm{B}}/v_{\mathrm{AB}}^2
\end{pmatrix},
\end{equation}
where the symbol \(\leftrightarrow\) unites the meaning of \(=\) and the meaning of \(\backsim\) while the symbol \(\odot\) unites the meaning of the usual product of two numbers and the meaning of the dot product of two spatial vectors; the symbol \(\otimes\) denotes the dyadic (outer) product.

Historically, the early attempts to obtain such a mapping did not lead to a correct/unambiguous expression of \(\mathbf{r}^{(\mathrm{B})}\) via \(\mathbf{r}^{(\mathrm{A})}\)  because of the failure to distinguish between a column vector and a true vector.\cite{early attempts} To keep being mathematically correct the later treatment\cite{Cushing1967} could not avoid resorting to coordinates but appeared limited to the transformation known as a boost (see the final remark in Section \ref{boosts}.)

\section{Coordinate transformations in the relativity theory}\label{coordinate transformations}

\subsection{Lorentz transformation}\label{Lorentz transformation}

As soon as one specifies coordinate systems in the spaces A and B, in terms of relations their unit base vectors \(\{\mathbf{e}^{(\mathrm{A})}_\alpha\}\) and \(\{\mathbf{e}^{(\mathrm{B})}_\alpha\}\) to some physical directions, one can arrive at the relationship
\begin{equation}\label{general transformation} 
\vec{\rho}^{(\mathrm{B})}=\mathbb{M}^{\mathrm{B}}_{\mathrm{A}}\vec{\rho}^{(\mathrm{A})}
\end{equation}
between the column vectors
\[
\vec{\rho}^{(\mathrm{A})}=
\begin{pmatrix}
ct^{(\mathrm{A})}\\
x^{(\mathrm{A})}\\
y^{(\mathrm{A})}\\
z^{(\mathrm{A})}
\end{pmatrix},
\quad
\vec{\rho}^{(\mathrm{B})}=
\begin{pmatrix}
ct^{(\mathrm{B})}\\
x^{(\mathrm{B})}\\
y^{(\mathrm{B})}\\
z^{(\mathrm{B})}
\end{pmatrix},
\]
made of time and spatial coordinates of a given event as observed in the inertial coordinate systems A and B, respectively.

With the choice
\[
\mathbf{e}^{(\mathrm{A})}_x=\frac{\mathbf{v}^{(\mathrm{A})}_{\mathrm{B}}}{ v_{\mathrm{AB}} },\quad \mathbf{e}^{(\mathrm{B})}_x=-\frac{\mathbf{v}^{(\mathrm{B})}_{\mathrm{A}}}{ v_{\mathrm{AB}} },\quad \mathbf{e}^{(\mathrm{A})}_y\backsim\mathbf{e}^{(\mathrm{B})}_y, \mathbf{e}^{(\mathrm{A})}_z\backsim\mathbf{e}^{(\mathrm{B})}_z
\]
for the unit base vectors Eqs.~\eqref{time relation AB}-\eqref{transversal relation AB} follow Eq.~\eqref{general transformation} with \(\mathbb{M}^{\mathrm{B}}_{\mathrm{A}}=\mathbb{L}(v_{\mathrm{AB}})\) where
\begin{equation}\label{Lorentz matrix} 
\mathbb{L}(v)=
\begin{pmatrix}
\gamma & -\gamma v/c & 0 & 0\\
-\gamma v/c & \gamma & 0 & 0\\
0 & 0 & 1 & 0\\
0 & 0 & 0 & 1
\end{pmatrix}
\end{equation}
is the matrix of the original form of Lorentz transformation.

\subsection{Boost transformations in the physics literature}\label{boosts}

Boosts make a well known class of the transformations \eqref{general transformation} introduced in graduate level physics courses with an aid of its matrix
\begin{equation}\label{boost transformation matrix}
\mathbb{M}^{\mathrm{B}}_{\mathrm{A}}=\mathbb{B}\left(\vec{v}^{(\mathrm{A})}_{\mathrm{B}}\right)
\end{equation}
where\cite{Boost matrix}
\begin{equation}\label{boost matrix}
\mathbb{B}(v\vec{n})=
\begin{pmatrix}
\gamma & -\gamma vn_x/c& -\gamma vn_y/c& -\gamma vn_z/c\\
-\gamma vn_x/c & 1+(\gamma-1)n_x^2 & (\gamma-1)n_xn_y & (\gamma-1)n_xn_z\\
-\gamma vn_y/c & (\gamma-1)n_yn_x & 1+(\gamma-1)n_y^2  & (\gamma-1)n_yn_z\\
-\gamma vn_z/c & (\gamma-1)n_zn_x & (\gamma-1)n_zn_y & 1+(\gamma-1)n_z^2 
\end{pmatrix}
,\quad \vec{n}=
\begin{pmatrix}
0\\
n_x\\
n_y\\
n_z
\end{pmatrix},
\end{equation}
\(n_x^2+n_y^2+n_z^2=1\), the elements of the column vector \(\vec{v}^{(\mathrm{A})}_{\mathrm{B}}\) are the A components of the velocity of B with respect to A.

Due to the identities\cite{Boost reduction}
\begin{equation}\label{boost reduction}
\mathbb{B}(v\vec{n})=\mathbb{R}^{-1}(\vec{n})\mathbb{L}(v)\mathbb{R}(\vec{n}),\quad \vec{e}_x=\mathbb{R}(\vec{n})\vec{n},
\end{equation}
where
\[
\vec{e}_x=\begin{pmatrix}
0\\
1\\
0\\
0 
\end{pmatrix},\quad
\vec{n}=\begin{pmatrix}
0\\
\sin\theta\cos\phi\\
\sin\theta\sin\phi\\
\cos\theta
\end{pmatrix},\quad 0\le\phi<2\pi, 0\le\theta<\pi,
\]
\[
\mathbb{R}(\vec{n})=
\begin{pmatrix}
1&0&0&0\\
0&\sin\theta\cos\phi&\sin\theta\sin\phi&\cos\theta\\
0 &-\sin\phi&\cos\phi&0\\
0 &-\cos\theta\cos\phi&-\cos\theta\sin\phi&\sin\theta
\end{pmatrix},
\] 
one can say that \(\mathbb{B}(\vec{v})\) incorporates \(\vec{v}\) as an observable direction.

It is easy to find that textbooks' authors exploit the boosts only to interpret the Thomas precession contribution to the spin motion of a relativistically moving charged particle.\cite{Boosts and Thomas precession} However, despite this seemingly definite connection to observable phenomena, all the texts up till now have favored the presentations that facilitate algebraic operations involving boosts over those which would treat a boost as a physically defined operation/motion or relations between physical objects.

Apparently, such a non-physical cannot lead to any better understanding of the spin precession or anything else because it leaves the physical meaning of \(\mathbb{B}(\vec{v})\) itself out of consideration. Its only benefit for physicists seems to be that it has found no reason to consider a boost as the Lorentz transformation ``without rotation,''  in attempt to generalize the idea of parallel transport, such as the definition (ii) at p.~871 in Ref.~\onlinecite{Frahm1979}. As a result, to ``a boost''  mathematicians  prefer to apply cautious terms such as ``an aligned axis Lorentz transformation'' (see p.~236 in Ref.~\onlinecite{Kennedy2002}.)

The most obvious manifestation of the problem is that no text attempts to formulate a physics based definition of a boost, thereby preventing any futher reasonable use of that concept. The reference formula \eqref{boost matrix} makes it difficult to get an idea how mutual orientations of several bodies change when set in motion, because \eqref{boost matrix} alone provides no hints how to choose the axes of the coordinate systems in use. The seemingly key formula \eqref{boost reduction} appears to be nothing but a relation between two representations \eqref{boost matrix} and \eqref{Lorentz matrix} for one relative motion, of which the formal simplicity of \eqref{Lorentz matrix} may even be misleading:

When asked about the direction of the \(x^{(\mathrm{A})}\)-axis, someone may correctly infer from \eqref{Lorentz matrix} that it is the direction of the \(x^{(\mathrm{B})}\)-axis instantaneously observed in the coordinate system A. But the problem is that a researcher must be able to identify any directions before establishing/verifying (theoretically/experimentally) the relationships \eqref{Lorentz matrix}. To avoid that apparent logical circle, one might take the above description of the direction as a definition and an explicit starting point in a derivation of the transformation rule \eqref{Lorentz matrix}. Needless to say that such an accurate approach can hardly be found in the existing, history-oriented, presentations of the relativity theory.

Now the coordinate-free description for the relation between two arbitrary spaces in Section \ref{transformation} allows one to formulate logically consistent and physically explicit definition of a boost.

\subsection{Definition of a boost and derivation of its matrix}

Aside from the origins, the above consideration refers to no elements of coordinate systems. To set up Cartesian coordinate systems in the spaces A and B one should specify the direction of their spatial base unit vectors \(\mathbf{e}_{\mathrm{A}\alpha}\) and \(\mathbf{e}_{\mathrm{B}\alpha}\) for \(\alpha=x,y,z\). Let
\begin{equation}\label{boost definition condition}
\mathbf{e}_{\mathrm{A}\alpha\perp}\backsim\mathbf{e}_{\mathrm{B}\alpha\perp}
\end{equation}
and
\begin{equation}\label{boost velocity components}
\left(\mathbf{e}_{\mathrm{A}\alpha}\cdot\mathbf{v}^{(\mathrm{A})}_{\mathrm{B}}\right)=-\left(\mathbf{e}_{\mathrm{B}\alpha}\cdot\mathbf{v}^{(\mathrm{B})}_{\mathrm{A}}\right)\equiv v_{\alpha}.
\end{equation}
In other words, in view of Eq.~\eqref{dot property}, \(-\mathbf{v}^{(\mathrm{B})}_{\mathrm{A}}\) makes the same angles with \(\mathbf{e}_{\mathrm{B}\alpha}\) in the space B as \(\mathbf{v}^{(\mathrm{A})}_{\mathrm{B}}\) does with \(\mathbf{e}_{\mathrm{A}\alpha}\) in the space A. By definition, a boost is the transformation \eqref{time transformation} of time along with the transformation of coordinates that satisfies the conditions \eqref{boost definition condition} and \eqref{boost velocity components}.

The equivalence relations \eqref{boost definition condition} and \eqref{transversal plane relation} along with the property \eqref{dot property} yield the equality of numbers
\begin{equation}\label{transversal equity}
\left(\mathbf{r}^{(\mathrm{A})}_{\perp}\cdot\mathbf{e}_{\mathrm{A}\alpha\perp}\right)=\left(\mathbf{r}^{(\mathrm{B})}_{\perp}\cdot\mathbf{e}_{\mathrm{B}\alpha\perp}\right).
\end{equation}
Eqs.~\eqref {boost velocity components} and \eqref{longitude transformation} allows one to rewrite the dot products as
\begin{multline*}
\left(\mathbf{r}^{(\mathrm{A})}_{\perp}\cdot\mathbf{e}_{\mathrm{A}\alpha\perp}\right)=\left(\mathbf{r}^{(\mathrm{A})}_{\perp}\cdot\mathbf{e}_{\mathrm{A}\alpha}\right)=
\left(\left(\mathbf{r}^{(\mathrm{A})}-r^{(\mathrm{A})}_{\parallel}\frac{\mathbf{v}^{(\mathrm{A})}_{\mathrm{B}}}{v}\right)\cdot\mathbf{e}_{\mathrm{A}\alpha}\right)=\\
r^{(\mathrm{A})}_{\alpha}-r^{(\mathrm{A})}_{\parallel}\frac{v_{\alpha}}{v}=r^{(\mathrm{A})}_{\alpha}-\frac{(\mathbf{v}^{(\mathrm{A})}_{\mathrm{B}}\cdot\mathbf{r}^{(\mathrm{A})})}{v}\frac{v_{\alpha}}{v}=r^{(\mathrm{A})}_{\alpha}-\frac{v_{\alpha}v_{\beta}}{v^2}r^{(\mathrm{A})}_{\beta}
\end{multline*}
and
\begin{multline*}
\left(\mathbf{r}^{(\mathrm{B})}_{\perp}\cdot\mathbf{e}_{\mathrm{B}\alpha\perp}\right)=\left(\mathbf{r}^{(\mathrm{B})}_{\perp}\cdot\mathbf{e}_{\mathrm{B}\alpha}\right)=
\left(\left(\mathbf{r}^{(\mathrm{B})}+r^{(\mathrm{B})}_{\parallel}\frac{\mathbf{v}^{(\mathrm{B})}_{\mathrm{A}}}{v}\right)\cdot\mathbf{e}_{\mathrm{B}\alpha}\right)=\\
r^{(\mathrm{B})}_{\alpha}-r^{(\mathrm{B})}_{\parallel}\frac{v_{\alpha}}{v}=r^{(\mathrm{B})}_{\alpha}-\gamma(r^{(\mathrm{A})}_{\parallel}-vt^{(\mathrm{A})})\frac{v_{\alpha}}{v}=\\
r^{(\mathrm{B})}_{\alpha}+\gamma v_{\alpha}t^{(\mathrm{A})}-\gamma\frac{(\mathbf{v}^{(\mathrm{A})}_{\mathrm{B}}\cdot\mathbf{r}^{(\mathrm{A})})}{v}\frac{v_{\alpha}}{v}=r^{(\mathrm{B})}_{\alpha}+\gamma v_{\alpha}t^{(\mathrm{A})}-\gamma\frac{v_{\alpha}v_{\beta}}{v^2}r^{(\mathrm{A})}_{\beta}
\end{multline*}
Here the common pithy notation is used: \(r_x\equiv x, r_y\equiv y, r_z\equiv z\) while the repeated index \(\beta\) implies the summation over all its values. Then Eq.~\eqref{transversal equity} entails
\[r^{(\mathrm{B})}_{\alpha}=r^{(\mathrm{A})}_{\alpha}-\gamma v_{\alpha}t^{(\mathrm{A})}+(\gamma-1)\frac{v_{\alpha}v_{\beta}}{v^2}r^{(\mathrm{A})}_{\beta},\]
which is just the spatial part of the transformation \eqref{general transformation} with the matrix \eqref{boost transformation matrix}.

In terms of column vectors the above equation can be written as
\begin{equation}\label{column vector boost transformation}
\vec{r}^{(\mathrm{B})}=\vec{r}^{(\mathrm{A})}-\gamma\vec{v}^{(\mathrm{A})}_{\mathrm{B}}t^{(\mathrm{A})}+(\gamma-1)\frac{\left( \vec{r}^{(\mathrm{A})}\cdot\vec{v}^{(\mathrm{A})}_{\mathrm{B}}\right)\vec{v}^{(\mathrm{A})}_{\mathrm{B}}}{v^2}.
\end{equation}
This form of a boost was obtained from Eq.~\eqref{boost reduction}  in Ref.~\onlinecite{Cushing1967}.

\subsection{Boost operation in the laboratory frame}\label{boost in the laboratory}

One can address the definition of a boost in the previous section so as to come to physically meaningful conclusions directly. An important example is a simple analytical description for the distortion that a Cartesian coordinate basis exhibits while simultaneously observed in the space where it experiences a boost operation.

\begin{figure}[h]
  \centering
 \includegraphics[width=10cm]{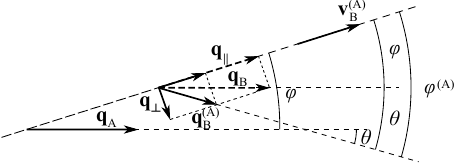}
  \caption{The moving vector \(\mathbf{q}_{\mathrm{B}}\) is instantaneously observed as the vector \(\mathbf{q}^{(\mathrm{A})}_{\mathrm{B}}\).}
   \label{moving vector}
\end{figure}

Fig.~\ref{moving vector} presents a plane parallel to a spatial vector \(\mathbf{q}_{\mathrm{A}}\) and the velocity \(\mathbf{v}^{(\mathrm{A})}_{\mathrm{B}}\)  of the space B, both the vectors being sets of simultaneous events in the space A. The boost operation applied to the vector
\[\mathbf{q}_{\mathrm{A}}=q_{\parallel}\frac{\mathbf{v}^{(\mathrm{A})}_{\mathrm{B}}}{v}+\mathbf{q}_{\mathrm{A}\perp}\]
results in the vector
\[\mathbf{q}_{\mathrm{B}}=-q_{\parallel}\frac{\mathbf{v}^{(\mathrm{B})}_{\mathrm{A}}}{v}+\mathbf{q}_{\mathrm{B}\perp},\quad \mathbf{q}_{\mathrm{B}\perp}\backsim\mathbf{q}_{\mathrm{A}\perp}\]
which is spatial in the frame B and parallel to the above plane, too. Due to the definition of the boost and the length contraction effect, in the frame A the vector \(\mathbf{q}_{\mathrm{B}}\) is perceived (instantaneously observed) as the vector
\[\mathbf{q}^{(\mathrm{A})}_{\mathrm{B}}=(q_{\parallel}/\gamma)\frac{\mathbf{v}^{(\mathrm{A})}_{\mathrm{B}}}{v}+\mathbf{q}_{\perp}.\]
Since
\[\tan\varphi=\frac{|\mathbf{q}_{\perp}|}{q_{\parallel}}\]
for the angle \(\varphi\) between \(\mathbf{q}_{\mathrm{A}}\) and \(\mathbf{v}^{(\mathrm{A})}_{\mathrm{B}}\) as well as between \(\mathbf{q}_{\mathrm{B}}\) and \(-\mathbf{v}^{(\mathrm{B})}_{\mathrm{A}}\), and
\[\tan\varphi^{(\mathrm{A})}=\gamma\frac{|\mathbf{q}_{\perp}|}{q_{\parallel}}\]
for the angle \(\varphi^{(\mathrm{A})}\) between \(\mathbf{q}^{(\mathrm{A})}_{\mathrm{B}}\) and \(\mathbf{v}^{(\mathrm{A})}_{\mathrm{B}}\), one can find that
\[\tan\theta=\frac{(\gamma-1)\tan\varphi}{1+\gamma\tan^2\varphi}\]
for the angle \(\theta=\varphi^{(\mathrm{A})}-\varphi\) between \(\mathbf{q}^{(\mathrm{A})}_{\mathrm{B}}\) and \(\mathbf{q}_{\mathrm{A}}\) (see Fig.~\ref{moving vector}.)

Sometimes it may be reasonable to change the laboratory coordinate system with the aid of a boost so as to reach simpler state of motion, to reduce the form of interaction law etc. Then, in the limit \(v\ll c\), the angle \(\theta=O(v^2/c^2)\) shows what relative error is not properly compensated in case one fails to transform all relevant quantities appropriately.

\section{Conclusion}

A boost direction for a pair of physical spaces is that spatial direction in one of the spaces along which the other space moves. Free motions of point particles make an instrumentation for identifying the boost direction as well as events on a straight line along that direction. The concept of a boost direction secures the formulation of the basic relativity effects in a physics-based manner, which, eventually, results in the relation between two spaces in terms of their position vectors and time moments for a given event.

Within the physics-based approach, addressing the transformation of coordinates implies specifying some coordinate systems in each space in terms of physical objects/directions. This yields a logically consistent and physically meaningful presentation of the coordinate transformations commonly exploited in the special relativity theory, which makes observable effects associated with those transformations evident. In particular, for a Cartesian coordinate system subjected to a boost, the coordinate-free technique of reasoning allows one to evaluate its instantaneously observable apparent distortion easily.


\begin{thebibliography} {5}

\bibitem{Einstein1905}A. Einstein ``Zur  Elektrodynamik der bewegter K\"orper,'' Ann. Phys. \textbf{17}, 891-921 (1905)

\bibitem{post-Einsteinian monographs}See, e.g., chs.~I-VI in Ref.~ \onlinecite{Silberstein1914}, chs.~I-IV in Ref.~ \onlinecite{Tolman1917}, chs.~I-IV in Ref.~ \onlinecite{Carmichael1920}, pt.~I in Ref.~ \onlinecite{Pauli1958}, chs.~IV-VI in Ref.~ \onlinecite{Born1962}.

\bibitem{Silberstein1914} L. Silberstein, \textit{The Theory of Relativity}, (Macmillan and Co., 1914)

\bibitem{Tolman1917}Richard C. Tolman, \textit{The Theory of Relativity of Motion}, (University of California Press, 1917)

\bibitem{Carmichael1920}Robert D. Carmichael, \textit{The Theory of Relativity}, 2nd edition (Wiley, 1920)

\bibitem{Pauli1958}W. Pauli, \textit{The Theory of Relativity}, (Pergamon Press, 1958)

\bibitem{Born1962}Max Born, \textit{Einstein's Theory of Relativity}, (Dover Publications, 1962) 

\bibitem{textbooks}See, e.g., chs.~I-IV in Ref.~ \onlinecite{Bergmann1976}, chs.~I-II in Ref.~ \onlinecite{Moeller1955}, ch.~1 in Ref.~ \onlinecite{Taylor1992}, Sections~11.1-11.4 in Ref.~ \onlinecite{Jackson1998}, Sections~7.1-7.3 in Ref.~ \onlinecite{Goldstein2000}.

\bibitem{Bergmann1976}Peter G. Bergmann, \textit{Introduction to the Theory of Relativity}, (Dover Publications, 1976)

\bibitem{Moeller1955}C. M{\o}ller, \textit{The Theory of Relativity}, (Oxford University Press, 1955)

\bibitem{Taylor1992}Edwin F. Taylor and John Archibald Wheeler, \textit{Spacetime Physics}, (W. H. Freeman and Co., 1992)

\bibitem{Jackson1998}John David Jackson, \textit{Classical Electrodynamics}, 3rd edition (Wiley, 1998)

\bibitem{Goldstein2000}Herbert Goldstein, Charles Poole, and John Safko, \textit{Classical Mechanics}, 3rd edition (Addison-Wesley, 2000)

\bibitem{Fock1964}V. Fock, \textit{The Theory of Space Time and Gravitation}, 2nd Revised Edition (Pergamon Press, 1964)

\bibitem{Hagedorn1964} R. Hagedorn , \textit{Relativistic Kinematics. A Guide to Kinematic Problems of High-energy Physics}, (W. A. Benjamin, Inc., 1964)

\bibitem{geometry due to electromagnetism or gravity} Since both electromagnetism and gravity manifest themselves via the concept of force, they allow one to distinguish directions and thereby support projective geometry at least. The next step should involve point particles, i.e. such entities that can strongly interact over a sufficiently short range. Then the application of first Newton's law to several particles on the same straight line helps one identify stationary particles, which can embody Euclidean points. It enables one to introduce the concept of length in the usual manner.

The discussed approach implies that the physical regularities involved can be presented in pre-Euclidean/prenumeric terms to describe experiments/observations more directly. To get an idea about a technique of such description, see Sections V and VI in Ref.~\onlinecite{prenumeric}.

\bibitem{non-inertial frame}There is no way to introduce a non-inertial reference frame consistent with Euclidean geometry, since the latter exists because of physical regularities that manifest themselves in an inertial reference frame. In other words, one can rotate a real rigid body, but it worsens the physical realization of Euclidean geometry by means of the relationships between parts of this body.

\bibitem{prenumeric} Serge A. Wagner, ``How to introduce physical quantities physically," \url{<http://arxiv.org/pdf/1506.04122v1>}

\bibitem{quantization} For example, if someone would choose to store information about a local direction in the angular momentum of a hydrogen atom, he will inevitably fail both due to the well known unavoidable quantization of that quantity and because of the small lifetime of the corresponding excited state of motion. In accordance to the simple consideration known as Bohr's theory of a hydrogen atom, it occurs over the spatial scale, called the Bohr radius and recognized as a characteristic atomic scale.

In general, the elementary charge \(e\), Plank's constant \(\hbar\), electron's rest mass \(m_e\), proton's rest mass \(m_p\) are well known magnitudes which reveal the quantization of electric charge, angular momentum and mass/energy, respectively. The dimensional and/or provisional theoretical analysis can yield a set of characteristic spatial scales
\[f\left(e^2/\hbar c, m_e/m_p, \dots\right)\hbar/m_ec\]
 where \(f(\alpha, \xi, \dots)\) is a dimensionless  function of the dimensionless quantities \(\alpha, \xi, \dots\) The cases \(f=\alpha\), \(f=1\) and \(f=\alpha^{-1}\) yield the well known scales: the classical electron radius (in fact, the characteristic size of proton/neutron), Compton wavelength of electron and the Bohr radius. Here the use of the CGS system of units is assumed, so one can take \(c\) as a parameter in Maxwell equations, not necessarily the observable speed of light.

Though no formal analysis of applicability of Euclidean geometry has yet been published, it is over scales less than Compton wavelength of electron where some authors have found it difficult to provide a consistent concept of a spatial coordinate. See p.~40 in Ref.~\onlinecite{Bjorken&Drell1964} and p.~2 in Ref.~\onlinecite{Berestetskii1982} for simple order-of-magnitude considerations and, e.g., Ref.~\onlinecite{Hegerfeldt1998} for the exemplars of existing theoretical approach to the problem.

\bibitem{Bjorken&Drell1964} James D. Bjorken, Sidney D. Drell, \textit{Relativistic Quantum Mechanics}, (McGraw-Hill, 1964)

\bibitem{Berestetskii1982}V. B.  Berestetskii, E. M. Lifshitz and L. P. Pitaevskii \textit{Quantum Electrodynamics. Course of Theoretical Physics, vol. 4}, 2nd edition, (Pergamon Press 1982)

\bibitem{Hegerfeldt1998} G. C. Hegerfeldt,  ``Instantaneous spreading and Einstein causality in quantum theory,'' Ann. d. Phys. \textbf{7}, 716-725 (1998)

\bibitem{first formulation of the relativity principle} The first more or less general formulation of the relativity principle is believed to belong H. Poincar\'e. See, e.g., p.~111 in Ref.~\onlinecite{Poincare1905} or p.~107 and p.~300 in Ref.~\onlinecite{Poincare1913}

\bibitem{Poincare1905} H. Poincar\'e, Science and Hypothesis (London and Newcastle-on-Cyne: The Walter Scott Publishing Co., 1905)

\bibitem{Poincare1913} H. Poincar\'e, The Foundation of Science (NY: The Science Press New York and Garrison, 1913)

\bibitem{early presentations} The thought that first Newton's law has the same form in all inertial frames was viewed as trivial and, for this reason, exploited implicitly (until the accurate presentation in Ref.~ \onlinecite{Fock1964} has appeared.) In contrast, Einstein's idea that electromagnetic phenomena, such as the propagation of a spherical wave, was apparently perceived as nontrivial. So the statement that a spherical wavefront keeps its form in all frames appeared to be a popular explicit premise for the derivation of Lorentz transformation; see, e.g., p.~100 in Ref.~ \onlinecite{Silberstein1914} and p.~9 in Ref.~ \onlinecite{Pauli1958}.

\bibitem{Fock presentation} This is just the condition in the equations (8.05) and (8.06) at p.~21 in Ref.~\onlinecite{Fock1964}. Interestingly, Fock starts his consideration with the general wavefront equation, which, in principle, can describe non-spherical wavefronts, e.g., from two interfering point sources. But eventually he has narrowed his inference with no explicit reasoning.

\bibitem{relativistic construction of geometry} See the flat spacetime limit for ``the remark" (e) at p.~69 followed by the reasoning at p.~82 in Ref.~\onlinecite{Ehlers1972} or the description of ``the second clock effect" at p.~122 in Ref.~\onlinecite{Ehlers1973}.

\bibitem{Ehlers1972} J. Ehlers, F. A. E. Pirani, A. Schild,  ``The Geometry of Free Fall and Light Propagation,'' in \textit{General relativity, papers honour of J. L. Synge}, edited by L. O'Raifeartaigh (Clarendon Press, Oxford, 1972), p.~63--84.

\bibitem{Ehlers1973} J. Ehlers, A. Schild,  ``Geometry in a Manifold with Projective Structure,'' Commun. Math. Phys. \textbf{32}, 119--146 (1973)

\bibitem{Fock and geometry} This is an impicit reason why Fock\cite{Fock1964} starts his inference of a boost with a generally nonlinear transformation  of the space and time variables and addresses some concepts of Riemannian geometry as elements of a suitable notation.


\bibitem{light clock}The first description of the light clock is given at Ref.~\onlinecite[p.~293]{Laue1917}.

\bibitem{Laue1917} Max von Laue, ``Die Nordstr\=omsche Gravitationstheorie. (Bericht.),'' Jahrbuch der Radioaktivit\=at and Elektronik \textbf{14}, 264-312 (1917)

\bibitem{early attempts} See the footnote 24 at pp. 10 and 11 in Ref.~\onlinecite{Pauli1958} and the solution to Problem 1-3 at pp. 8 and 9 in Ref.~\onlinecite{Hagedorn1964} for the examples of such reasoning in educational texts for physicists.

\bibitem{Cushing1967} James T. Cushing, ``Vector Lorentz Transformations,'' Am. J. Phys. \textbf{35}, 858-862 (1967)

\bibitem{Boosts and Thomas precession} Cf. Section 7.3 in Ref.~\onlinecite{Goldstein2000} , Sections 11.7 and 11.8 in Ref.~\onlinecite{Jackson1998}, \S 41.4 in Ref.~\onlinecite{Misner1973}.

\bibitem{Misner1973} Charles W. Misner, Kip S. Thorne, John Archibald Wheeler, \textit{Gravitetion},  (Freeman, 1973)

\bibitem{Boost matrix} See, e.g., Eq. (11.98) at p.~547 in Ref.~\onlinecite{Jackson1998} or Eq. (24) at p.~872 in Ref.~\onlinecite{Frahm1979}.

\bibitem{Frahm1979} Charles P. Frahm, ``Representing arbitrary boosts for undergraduates,'' Am. J. Phys. \textbf{47} (10), 870-872 (1979)

\bibitem{Boost reduction} See, e.g., Eq.~(13) in Ref.~\onlinecite{Cushing1967}.

\bibitem{Kennedy2002} W. L. Kennedy, ``Thomas rotation: a Lorentz matrix approach,'' Eur. J. Phys. \textbf{23}, 235-247 (2002)
\end{thebibliography}
\end{document}